\documentclass[namedreferences]{solarphysics}

\usepackage[hyperref,optionalrh,showbiblabels]{spr-sola-addons} 
\usepackage[hyperref,optionalrh]{spr-sola-addons} 
\usepackage[optionalrh]{spr-sola-addons} 
\usepackage{graphicx}        
\usepackage{color}           
\usepackage{breakurl}        

\chardef\us=`\_
\definecolor{orange}{rgb}{1,0.5,0}
\begin{document}

\begin{article}
\begin{opening}

\title{Indirect Solar Wind Measurements Using Archival Cometary Tail Observations}

\author[addressref={aff1},email={ned@geo.phys.spbu.ru}]
{\inits{N.V.}\fnm{N.}~\lnm{Zolotova}}
\address[id=aff1]{St. Petersburg State University, Universitetskaya nab. 7/9, 198504 St. Petersburg, Russia}

\author[addressref=aff2]{\inits{Yu.~V.}\fnm{Yu.}~\lnm{Sizonenko}}
\address[id=aff2]{Main Astronomical Observatory, National Academy of Sciences of Ukraine, Akademika Zabolotnoho ul. 27, 03143 Kyiv, Ukraine}

\author[addressref={aff1}]
{\inits{M.V.}\fnm{M.}~\lnm{Vokhmyanin}}

\author[addressref={aff3,aff4,aff5}]
{\inits{I.S.}\fnm{I.}~\lnm{Veselovsky}}
\address[id=aff3]{Lomonosov Moscow State University, GSP-1, Leninskie Gory, 119991 Moscow, Russia}
\address[id=aff4]{Skobeltsyn Institute of Nuclear Physics, Lomonosov Moscow State University, GSP-1, Leninskie Gory, 119991 Moscow, Russia}
\address[id=aff5]{Space Research Institute, Russian Academy of Sciences, Profsoyuznaya ul. 84/32, 117997 Moscow, Russia}


\begin{abstract}

The paper addressed to the problem of the solar wind behaviour during the Maunder Minimum. Records on plasma tails of comets would allow to shed light on the physical parameters of the solar wind in the past. We analyse descriptions and drawings of comets to the eighteenth century. To differentiate dust and plasma tails, we address to their color, shape and orientation. Basing on the calculations made by F.A.~Bredikhin, we found that deviation of cometary tails from the antisolar direction on average exceeded 10$^{\circ}$, that is typical for dust tails. Catalogues of Hevelius and Lubieniecki are also examined. The first indication of plasma tail was revealed only for Great comet C/1769~P1.

\end{abstract}
\end{opening}

\section{Introduction}
     \label{S-Introduction} 

Apart from spacecraft measurements, the plasma tails are natural probes of the interplanetary space. Physical parameters of the solar wind during the so-called Grand minima are still the outstanding puzzle \textcolor{orange}{\citet{1976Sci...192.1189E}}. For the Maunder minimum (thereafter MM) from 1645 to 1715, \textcolor{orange}{\citet{1976IAUS...71....3P}} conjectured that whether a fast and persistent solar wind emanated from a coronal hole, entirely covering the Sun, or whether the solar wind just did not blow.

\textcolor{orange}{\citet{1836AnP...114..498B}} suggested that cometary particles move under the action of the effective repulsive force. This idea formed the basis of the mechanical theory of cometary shapes. \textcolor{orange}{\citet{Bredichin1879a,Bredichin1879b,Bredichin1880,Bredichin1886}} revised Bessel's calculations and created a well-known classification of cometary tails. This classification was further developed by \textcolor{orange}{\citet{Orlov1935}} and \textcolor{orange}{\citet{Wurm1954}}. However, the mechanical theory could not explain the large acceleration of cloud masses observed in type~I tails. \textcolor{orange}{\citet{1951ZA.....29..274B}} showed that the acceleration of the ionized particles in cometary tails can not be explained by the repulsive force of light pressure. He suggested that the ions are accelerated by the flow of the electrically charged particles emanating from the Sun, referred to as the solar wind, whose mathematical theory was developed by \textcolor{orange}{\citet{1958ApJ...128..664P}}.

\textcolor{orange}{\citet{1979P&SS...27.1001S}} proposed that during the MM the solar wind speed or the interplanetary magnetic field or both were low and not irregular. \textcolor{orange}{\citet{1997AnGeo..15..397M}} evaluated average velocities of the solar wind using proxy $aa$ index and obtained 194.3\,km\,s$^{-1}$ from 1657 to 1700 and 218.7\,km\,s$^{-1}$ from 1700 until now. \textcolor{orange}{\citet{1998GeoRL..25..897C}} extrapolated observed solar wind variations to the MM conditions and suggested that average velocities have an upper limit of 340\,$\pm$\,50\,km\,s$^{-1}$, and that of the interplanetary magnetic field, 0.3\,$\pm$\,0.1\,nT. Idea of the lowest conceivable solar wind magnetic field (the floor) was developed by \textcolor{orange}{\citet{2007ApJ...661L.203S}} and \textcolor{orange}{\citet{2012IAUS..286..179C}}.

The scenario of a giant coronal hole over the entire solar disk at the MM might imply the solar wind speed 800\,km\,s$^{-1}$ \textcolor{orange}{\citep{2010JGRA..115.1104S}}. \textcolor{orange}{\citet{2013ApJ...764...90W}} simulated the axial solar dipole during grand minima and concluded that a factor of 4\,--\,7 decrease in the total open flux should result in a similar suppression in the solar wind densities not affecting the solar wind speeds. Using geomagnetic indices as a proxy for the solar wind speed, \textcolor{orange}{\citet{2014ApJ...781L...7L}} conjectured that solar wind speeds would be relatively uniform in the MM (between 250 and 275\,km\,s$^{-1}$). \textcolor{orange}{\citet{2015ApJ...802..105R}} argued that disappearance of the solar wind during the MM is not compatible with the presence of geomagnetic activity and aurora which did not cease. \textcolor{orange}{\citet{2017NatSR...741548O}} noted that during the MM solar maximum regime prevailed with short-living fast wind at high latitudes. Results agree with the MM coronal magnetic field configuration constructed in a global MHD model by \textcolor{orange}{\citet{2015ApJ...802..105R}}.

\textcolor{orange}{\citet{2015ARep...59..791G}} analysed historic drawings of comets using the mechanical classification of cometary tails. He concluded that straight type~I tails depicted by observers in the MM correspond to the plasma tails, which in turn indicates the solar wind is a persistent phenomenon.

In this article, we are raising the issue whether observers had reported the plasma tails of comets in the distant past. In Section~\ref{S-Parameters}, the main parameters of a cometary tail which may help specify its nature (dust or plasma) are mentioned. In Section~\ref{S-Results}, we analyse color, shape, and orientation of the cometary tails from historic archives. Catalogues of Hevelius and Lubieniecki are also discussed. Section~\ref{S-Conclusions} accumulates conclusions.

\section{Parameters of Cometary Tails} \label{S-Parameters}

\begin{figure}     
  \centering
 \includegraphics[scale=0.5]{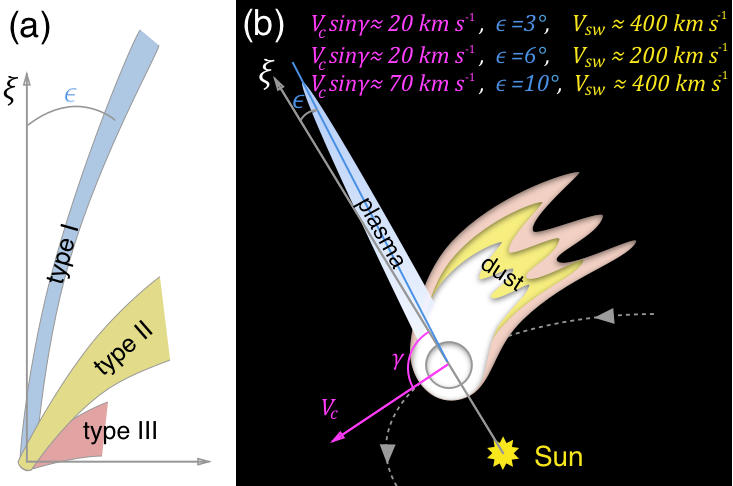}
 \caption{(a) Mechanical classification of cometary tails according to \citet{Bredichin1879b}. $\xi$ is the prolonged radius-vector. $\epsilon$ is the angle between the observed tail and the antisolar direction. Colors denotes type of tail. (b) Schematic illustration of the transit of a comet in the Solar System.}
 \label{Fig1}
\end{figure}

In mechanical theory, the ratio of the outward radial force $\mathbfit F_{\rm  r}$ to the gravitational force $\mathbfit F_{\rm g}$ given $|\mathbfit F_{\rm  r}/\mathbfit F_{\rm  g}|=R$ (the effective repulsive force) is used to divide the cometary tails into three types. Type~I tails have $R>12$; type~II, $2.2<R<0.9$; and type~III, $0.75<R<0.2$ \textcolor{orange}{\citep{Bredichin1879b}}. In different editions, these values slightly vary. Type~I tails are narrow, nearly straight and point almost directly away from the Sun. Type~II tails are smooth, curved and distant from the prolonged radius-vector of a comet. Type~III tails are short and broad.

Figure\textcolor{orange}{~\ref{Fig1}a} shows the classification of the tails according to \textcolor{orange}{\citet{Bredichin1879b}}. Blue color denotes type I tail which consists of hydrogen. Yellow color denotes type II tail which is composed of medium-heavy elements. Pink color corresponds to heavy metals of type III tail. $\xi$ is the prolonged radius-vector. $\epsilon$ is the angle between the observed tail and the antisolar direction. The lighter the chemical element, the closer its path to the prolonged radius-vector, and the greater $R$ for this element. Here, we emphasize that this classification is no longer valid.

The modern theory distinguish two basically different types of tails which are composed of either plasma or dust. Figure\textcolor{orange}{~\ref{Fig1}b} schematically illustrates the transit of a comet in the Solar System. Yellow decagon depicts the Sun. Dashed curve with arrows denotes a cometary orbit. A white circle is a coma complemented by a wide and curving dust tail and a thin and straight plasma tail. $\mathbfit V_{\rm c}$ is the comet's orbital speed; $\gamma$ is the angle between the cometary speed $\mathbfit V_{\rm c}$ and antisolar vector $\xi$. Simple calculations give that deviation of the plasma tail from the prolonged radius vector is defined as $\epsilon \approx \arctan(\mathbfit V_{\rm c} \sin \gamma / \mathbfit V_{\rm sw})$, where $\mathbfit V_{\rm sw}$ is the radial component of the solar wind flow \textcolor{orange}{\citep{2007hste.book..494M}}. Figure\textcolor{orange}{~\ref{Fig1}b} shows values of $\epsilon$ and $\mathbfit V_{\rm sw}$ for typical value of the transverse component of the comet's orbital speed $\mathbfit V_{\rm c} \sin \gamma \approx 20$\,km\,s$^{-1}$. Due to the retrograde orbit, Haley's comet has one of the highest velocities $\mathbfit V_{\rm c} \approx 70$\,km\,s$^{-1}$. Sophisticated calculations yield that typical value of $\epsilon<6^{\circ}$ for the plasma tails \textcolor{orange}{\citep{1967ApJ...147..201B,2007hste.book..494M}}. Deviation of dust tail from the antisolar direction amounts to first tens of degrees. For historic observations, magnitude of $\epsilon$ was restored by Bessel and revised by Bredikhin. In order to specify the nature of the tails reported by observers in the distant past, we will rely on this parameter.

Another feature of the dust tail is its curvature. In other words, if observer write that cometary tail became curved, this denotes dust tail. However, in the case of non-stationary outflow of cometary particles and at a short distance from a nucleus, the dust tail may appear straight. For a visual observer to discover the curvature of a tail, its length should be at least 6\,--\,7$^{\circ}$.

Next circumstance that has to be taken into account is the projection effect. The closer the Earth to the comet's orbital plane, the straighter the dust tail.

Finally, a bright comet often exhibits white dust tail which may assume a yellowish or reddish tint, accompanying the dim bluish plasma tail. For a naked-eye observer, dust tail shining by reflected sunlight is much brighter than that of plasma whose light is caused by the fluorescence of ionized gases. Their bluish radiation is located on the edge of the visual spectrum, to which the human eye is poorly sensitive. Therefore, historic reports on bright light emanating from cometary head also indicate the dust tail.

\section{Results} \label{S-Results}

\subsection{Color of Cometary Tail} \label{S-Color}
We analyse text descriptions of comets from the eleventh to eighteenth century. The older reports often mention the color of a comet. To find observations of bluish cometary tails, we address to \textcolor{orange}{\citet{Svyatsky2007}}, who collected astronomical events from the Complete collection of Russian chronicles; \textcolor{orange}{\citet{1871obco.book.....W}}, who translated the Chinese annals; catalogues of \textcolor{orange}{\citet{1984cdc..book.....K,1999ccc..book.....K}}, and other numerous reports of individual European observers.

In 11 volumes of the Complete collection of Russian chronicles, we did not find a single mention of the bluish tint of comets. On the contrary, red color of comets is often mentioned, usually as omen of bloodshed. Note, that a significant portion of comets was observed at sunset or at sunrise when the sky is painted with orange and red light. In European reports, we also did not come upon the bluish color of tails. In Eastern chronicles (Chinese, Korean, and Japanese), luminous envelopes and tails of comets are usually reported as white light objects. We would like to point out two translations of the report on comet C/1462~M1 from the Chinese annals. \textcolor{orange}{\citet{1871obco.book.....W}} wrote that colour of a star (comet) was a bluish white, while \textcolor{orange}{\citet{1984cdc..book.....K}}, that color was darkish white.

To conclude, we did not find robust support of hypothesis that observers saw bluish tails. Below we list comets whose observations were detailed.

\subsection{Great Comet C/1471 Y1} \label{sec:Great1471}

The comet was reported by Toscanelli \textcolor{orange}{\citep{Uzielli1893}}, who drew comet with a short tail. \textcolor{orange}{\citet{Bredichin1886}} defined $R=6.5$ (he proposed type I). Perihelion of the comet is on March 1 1472. On January 20 1472 $\epsilon=6^{\circ}$, and February 2, $\epsilon=18^{\circ}$, that indicates the dust tail. However, the comet was near the ecliptic plane, therefore the projection effects were significant, that results in large errors of calculations.

\subsection{Great Comet C/1577~V1} \label{sec:Great1577}

\begin{figure}   
 \centering
 \includegraphics[scale=0.24]{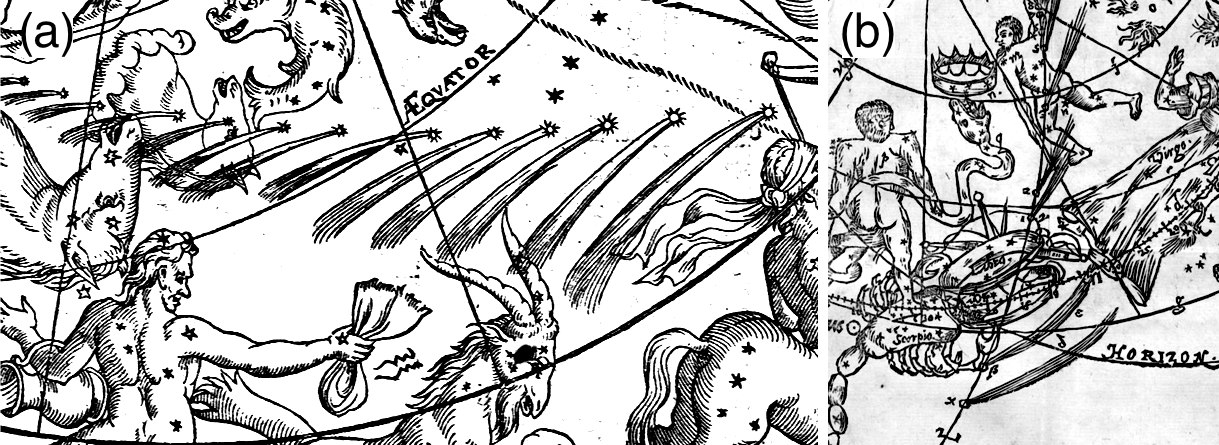}
 \caption{(\textbf{a}) Drawing of Great comet C/1577~V1 by \citet{Hayck1578}. (\textbf{b}) Fragment of drawing of Great comet C/1618~W1 by \citet{Montlhery1619}.}
 \label{Fig2}
\end{figure}

On November 23 1577, the cometary tail was pointed at evening. The head of the comet was above the front of the Equiculi constellation (modern Equuleus). The tail was curved, the convex part at Zenith. The tail ends in the front of the Pegasi constellation (modern Pegasus). The head of the comet was white, but not as bright as the light of stars. The tail was dim and reddish, especially near the head, like a flame breaking through the smoke, as we translate it from \citet[``Caudam porrigebat hoc vespere, in eam Stellulam, qu{\ae} est superior in fronte Equiculi... Erat autem, \`{a} capite versus dictam Stellam, paulum more solito incuruata, conuexam partem in Zenith tollens, ade\`{o}, ut si \`{a} capite per dictam Stellam ulterius protrahi fingeretur, suo ductu obliquo 
versus eam pertingeret, qu{\ae} est in fronte Pegasi. Color autem capitus Comet{\ae} suit albus, non tam clarus, sed pallidior, neque ita lucidus, ut Stellarum lumen. Cauda ver\`{o} obscuram rubedinem, praesertim quo erat capiti vicinior, ostendebat, qualis fer\`{e} solet esse flamm{\ae} alicujus, per fumum densum eluctantis...'']{brahe1610}. Numerous engravings by \textcolor{orange}{\citet[see here Figure~\ref{Fig2}a]{Hayck1578}}, \textcolor{orange}{\citet{Bazelio1578,Busch1577,Graminaeum1578,Maestlino1578}}, and others describe a long and curved cometary tail, which corresponds to the dust nature of the tail.

\textcolor{orange}{\citet{Bredichin1886}} also defined that the comet had the dust tail with $0.02<R<2.8$. From November 13 1577 to January 12 1578, the angle of deviation of the asymptote of the hyperbola of the cometary tail from the antisolar direction varied from 12.5$^{\circ}$ to 33.5$^{\circ}$ and on average was 20$^{\circ}$ \textcolor{orange}{\citep{Bredichin1879a}}. Here, we may conclude that observers reported the dust tail.

\subsection{Comet C/1580 T1} \label{sec:Comet1580}

The comet was reported by \textcolor{orange}{\citet{Graminaeus1581}}. \textcolor{orange}{\citet{Bredichin1886}} defined $R=1$ and $\epsilon=15^{\circ}$ that indicates the dust tail. However, since there is only one observation of the tail, the inaccuracy of the estimates is large \textcolor{orange}{\citep{Bredikhin1862}}.

\subsection{Great Comet C/1582 J1} \label{sec:Great1582}

Basing on three observations by Tycho Brahe on May 12, 17, and 18 1582, \textcolor{orange}{\citet{Bredichin1879b,Bredichin1886}} defined $R=0.2$ and $\epsilon = 37.5^{\circ}$ (dust tail). Note that the orbital elements are uncertain, due to lack of observations.

\subsection{Great Comet C/1618 W1} \label{sec:Great1618}

There are numerous drawings of this comet by \textcolor{orange}{\citet[see here Figure~\ref{Fig2}b]{Bainbridge1619,Cysato1619,Keplero1619,Montlhery1619}}, and \textit{etc}. \textcolor{orange}{\citet{Bredichin1886}} calculated $0.6<R<2.1$. From November 30 1618 to January 16 1619, $\epsilon$ varied from 16$^{\circ}$ to 41.6$^{\circ}$ and on average was 25$^{\circ}$ \textcolor{orange}{\citep{Bredichin1879a}}, indicating the dust tail.

\subsection{Comet C/1652 Y1} \label{sec:Comet1652}

\begin{figure}  
  \centering
 \includegraphics[scale=0.35]{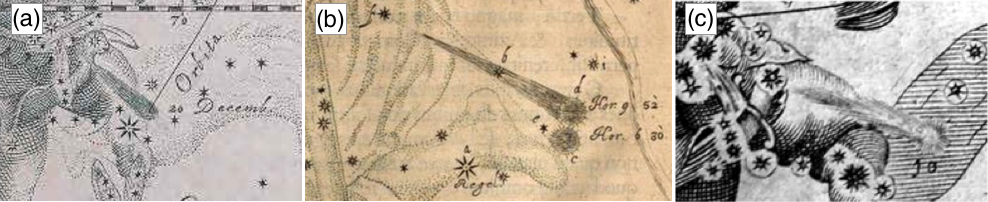}
 \caption{Fragments of drawings of comet C/1652 Y1 by \citet{1668ctnc.book.....H} (\textbf{a}) and (\textbf{b}), and \citet{Weigel1653} (\textbf{c}).}
 \label{Fig3}
\end{figure}

The comet was observed more than a month after the perihelion. According to the reports of \textcolor{orange}{\citet{1668ctnc.book.....H}} and \textcolor{orange}{\citet{Weigel1653}}, on December 20 1652 (December 10, Old Style) the comet passed through the foot of Orion and its tail had the largest size among the entire set of its observations (Figure\textcolor{orange}{~\ref{Fig3}}). Hevelius drew a straight tail 5$^{\circ}$ in length, and Weigel, slightly curving tail 6$^{\circ}$ in length. Recall that a short tail (less than 6\,--\,7$^{\circ}$) seems for a visual observer almost straight.

Of particular interest is the dynamics of the cometary tail in Figure\textcolor{orange}{~\ref{Fig3}b}. \textcolor{orange}{\citet{1668ctnc.book.....H}} described observations as follows. At half past six, the comet was visible in position $c$, tail passed through the star $e$ and star $b$ which is above the foot of Orion in Eridan. Moreover, Regel (moderm Rigel), star $b$, and the comet form an almost equilateral triangle... Distance between the comet and the star $b$ was 3$^{\circ}$\,25''... At 9 hours 52 minutes, the comet was already slightly higher at position $d$. The tail was still located along the foot of Orion in Eridan and pointed toward the star $f$ in the hilt of the sword... ``Hor\^a dimidi\^a  circiter septim\^a, Cometa in \textit{c} prim\`um nobis illuxit, caudam extendens per \textit{e} Stellulam ... \& Stellulam \textit{b}, supra pedem Orionis in Eridano, \textit{b} supremam cinguli circiter vers\`us. At\'q; tum Trianqulum fer\`e \ae quilaterum, cum Regel, \& dict\^a Stell\^a supra pedem Orionis constituebat... Distantia autem... inter Cometam \& stellam \textit{b}, 3~grad. 25~min... At hor\^a 9 52\,$'$, Cometa jam paul\`o altiorem occupaverat locum, nempe \textit{d}... Cauda ver\`o adhuc per stellam pedis Oriones in Eridano, us\'q; fer\`e stellam \textit{f} in Ensis manubrio...''). Thus, in three and a half hours the position angle of the tail has changed by 18$^{\circ}$. Such dramatic changes are hardly probable even under the influence of the local inhomogeneity of the solar wind.

\textcolor{orange}{\citet{Bredichin1879b}} defined average $R=1.5$. On December~20, $\epsilon=15^{\circ} 15'$; December~23, $\epsilon=13^{\circ}$; and December~26, $\epsilon=7^{\circ}$. As only three observations were available, the uncertainty is high. However, all of the above findings indicate that observers reported the dust tail.

\subsection{Comet C/1661 C1} \label{sec:Comet1661}

\begin{figure} 
 \centering
 \includegraphics[scale=0.28]{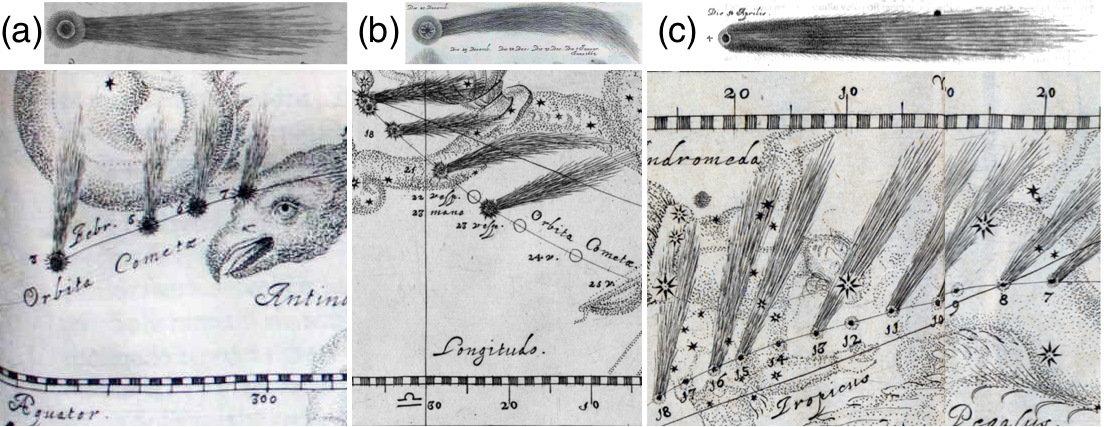}
 \caption{Fragments of drawings of comet C/1661 C1 by \citet{1668ctnc.book.....H} (\textbf{a}), Great comet C/1664~W1 by \citet{Hevelii1665} (\textbf{b}), Great comet C/1665 F1 by \citet{Hevelii1666} (\textbf{c}).}
 \label{Fig4}
\end{figure}

The comet was observed by \textcolor{orange}{\citet{1668ctnc.book.....H,Megerlino1661,Welper1661}}, and others. Welper drew a slightly curved tail. According to Hevelius (Figure\textcolor{orange}{~\ref{Fig4}a}), the maximal length of the tail was 5\,--\,6$^{\circ}$ on February 3 1661. Such short tails seem for a visual observer almost straight. \textcolor{orange}{\citet{Bredikhin1862}} noted poor quality of the observations.

\subsection{Great Comet C/1664 W1} \label{sec:Great1664}

This comet was observed by \textcolor{orange}{\citet{Cassini1665,Montanari1665}}, and others. According to \textcolor{orange}{\citet{Hevelii1665}} on December 21 1664, the comet had the longest and curving tail (Figure\textcolor{orange}{~\ref{Fig4}b}). \textcolor{orange}{\citet{Bredichin1879b,Bredichin1886}} evaluated $R \approx 1.2$\,--\,1.7 and $\epsilon = 27^{\circ}$, which corresponds to the dust tail. Also note that at the beginning of the observations the Earth passed through the plane of the cometary orbit, and then the full moon interfered with the observations. The cometary orbit was inclined by 21$^{\circ}$ to the ecliptic, therefore the projection effects might be significant. Therefore, the quality of the observations is poor, and the value of $\epsilon$ is inaccurate.

\subsection{Great Comet C/1665 F1} \label{sec:Great1665}

The comet was observed in the morning twilight, not far from the horizon. \textcolor{orange}{\citet{Hevelii1666}} drew a wide, straight tail (Figure\textcolor{orange}{~\ref{Fig4}c}). \textcolor{orange}{\citet{Lubienietski1681}} also showed a schematic comet with a straight tail. However, notice that Lubieniecki made his drawing from the words of Thomas Bartholin, who observed in Copenhagen.

\textcolor{orange}{\citet{Bredichin1879a,Bredichin1886}} wrote that quality of the observations is poor. $R$ changes from 2.9 to 115.8; $\epsilon$ from $-2$ to 12.9. \textcolor{orange}{\citet{Bredikhin1862}} concluded that large variation of $R$ raises doubts that the comet had type~I tail.

\subsection{Great Comet C/1668 E1} \label{sec:Great1668}

\textcolor{orange}{\citet{Cassini1730}} reported this observation as meteorological phenomenon similar to that of 1683. In 1702, \textcolor{orange}{\citet{Maraldi1743}} also reported another similar phenomenon at the same place. Observational data are poor to define the nature of this object.

\subsection{Comets C/1672 E1 and C/1677 H1} \label{sec:Comets16721677}

Comet C/1672~E1 was drawn by \textcolor{orange}{\citet{Hevelius1672}} and \textcolor{orange}{\citet{Weigel1672}}, and comet C/1677~H1, by \textcolor{orange}{\citet{Honold1677,Hooke1678,Voigt1677}}. The tails of comets were too short to defined their nature.

\subsection{Great Comet C/1680 V1} \label{sec:Great1680}

\begin{figure} 
 \centering
 \includegraphics[scale=0.35]{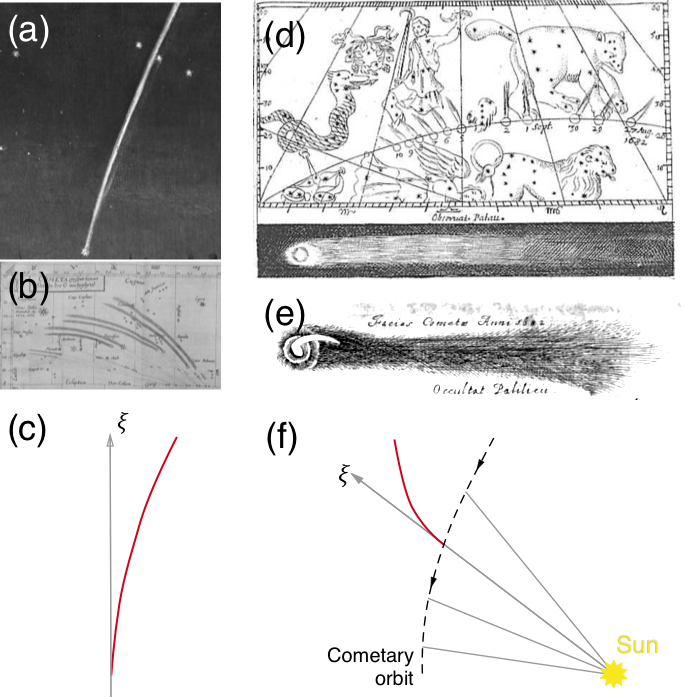}
\caption{Fragments of (\textbf{a}) the painting by Lieve Verschuier, (\textbf{b}) drawing by \citet{Voigt1681} of Great comet C/1680~V1. (\textbf{c}) Deviation of the cometary tail from the prolonged radius-vector $\xi$ sketched from \citet{Bredichin1880}. Fragments of drawing of comet 1P/1682 (\textbf{d}) by \citet{Montanari1682} and (\textbf{e}) by \citet{Hevelii1685}. (\textbf{f}) Deviation of the cometary tail from the prolonged radius-vector sketched from \citet{Bredichin1880}.}
 \label{Fig5}
\end{figure}

The first Astronomer Royal John Flamsteed reported the comet from December 10 1680 to February 5 1681. On the first day, he noted that the cometary tail was thin and stood straight up from the Horizon, as translated by us from \textcolor{orange}{Flamsteed (\citeyear{Flamsteed1725}}, ``Comet{\ae} Caudam pr{\ae}tenuem advertimus, ab Horizonte sursum erectam...''). On the next day, the tail was like a beam standing straight up from the Horizon and slightly deviating to the right of the vertical (``Cauda quasi trabs ab Horizonte sursum erecta conspiciebatur, ad dextram etiam parum a perpendiculo declinaris...''). Note that ``trabs'' could be translated as timber, beam, rafter, or tree trunk. On December 21 (perihelion on December 18), the tail was located to the right of the Dolphin constellation, slightly curving (``Cauda fixas Delphini \`{a} dextra reliquerat, incurvata paulul\`{u}m...''). On December 26, the tail was thinner and shorter than before (``Cauda rarior quam antehac \& brevior...''). On January 13 1681, the tail was very weak, but sufficiently wide (``Cauda valde debilis satis tamen lata fuit'').

The comet had small perihelion distance ($q=0.006$~AU) and was observed overall in Europe. There are numerous engravings of this comet with straight \textcolor{orange}{\citep{Mayern1681,Cassini1681}} or curved \textcolor{orange}{\citep[Figure~\ref{Fig5}]{Merian1691,Voigt1681}} tail. 

Figure\textcolor{orange}{~\ref{Fig5}a} shows a fragment of the painting by Lieve Verschuier ``The Great Comet of 1680 over Rotterdam''. To the right of bright tail, another dim tail is depicted. Plasma tails well known to be always located on the leading edge of dust tails. One can hypothesize that bright tail is a syndyne, and dim tail, a synchrone. Syndyne is the locus of the dust particles for a given radiation pressure and synchrone being the locus of particles emitted at the same time.

For the bright tail, \textcolor{orange}{\citet{Bredichin1880,Bredichin1886}} defined $R=1$ and $\epsilon$ varied from 1$^{\circ}$ to 20$^{\circ}$ (generally $\epsilon>10^{\circ}$, dust tail). Figure\textcolor{orange}{~\ref{Fig5}c} shows the hyperbola asymptote (red curve) of the cometary tail and the prolonged radius-vector $\xi$. Figure is sketched from \textcolor{orange}{\citet{Bredichin1880}}.

\subsection{Halley's Comet 1P/1682 Q1} \label{sec:Halley1682}

The comet (Figure~\textcolor{orange}{\ref{Fig6}}) was observed by \textcolor{orange}{\citet{Hevelii1685,Honold1682,Kirch1682,Montanari1682}}, and others. Hevelius (Figure~\textcolor{orange}{\ref{Fig5}e}) drew a bright, short, and curving jet emanating from the nucleus: on September 8, in the evening, the comet's head was visible in the optical tube. The nucleus not only kept its oval shape. A bright curving jet (or ray) prolonged from the nucleus to the tail \textcolor{orange}{Hevelius (\citeyear{Hevelii1685}}, ``Die 8 Sept. vesp... Caput Cometae h\^{a}c die per Tubum Opticum merebatur videri; non sol\`{u}m qu\`{o}d constanter clarissimum nucleum figur{\ae} ovalis conservaret, sed simul incurvatum splendidissimum radium, ad ipso nucleo in Caudam usque, sese extendentem exhiberet.'').

According to \textcolor{orange}{\citet{Bredichin1880}}, the orientation of the Halley's cometary tail 1P/1682~Q1  corresponds to type I with $R=12$ and $\epsilon=23^{\circ}$ (Figure~\textcolor{orange}{\ref{Fig5}f}. Black arrows mark the comet's path, $\xi$ is the prolonged radius-vector, and red curve is the asymptote of the tail. The value of $\epsilon$ indicates the dust nature of the Halley's cometary tail.

\subsection{Comets C/1739 K1 and C/1742~C1} \label{sec:Comets17391742}

Comet C/1739~K1 was drawn by \textcolor{orange}{\citet{Zanotti1739}}, and comet C/1742~C1, by \textcolor{orange}{\citet{Zanotti1742}} and \textcolor{orange}{\citet{Wiedeburg1742}}. The tails of both comets are too short to define their nature.

\subsection{Great Comet C/1743 X1} \label{sec:Great1743}

\begin{figure} 
 \centering
 \includegraphics[scale=0.23]{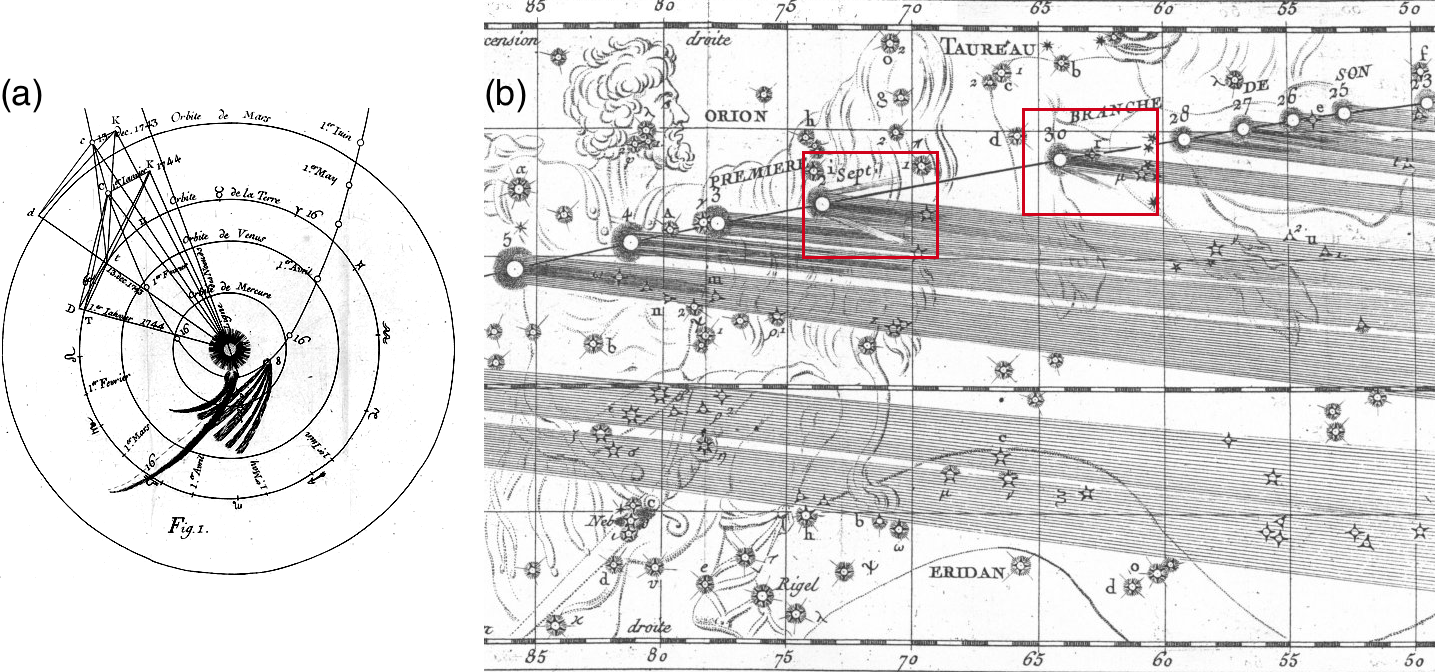}
 \caption{Drawing of Great comet C/1743~X1 by \citet{Cheseaux1744} (\textbf{a}). Fragment of drawing of Great comet C/1769~P1 by \citet{Messier1778} (\textbf{b}). Red blocks mark the nights when cometary tail had a complex structure.}
 \label{Fig6}
\end{figure}

The comet had quite complex structure. Few days after the perihelion, numerous observers \textcolor{orange}{\citep{Cheseaux1744,Euler1744,Heinsius1744,Zanotti1744}} reported multiple tail. Figure~\textcolor{orange}{\ref{Fig6}a} shows that the cometary tail is curved and significantly delayed relative to the prolonged radius-vector, that is typical for dust tails.

\textcolor{orange}{\citet{Bredichin1880,Bredichin1886}} quantified $R=1; 2.4; 12;$ and 17; he concluded that the multiple tail belongs to type I and II.

\subsection{Great Comet C/1769 P1} \label{sec:Great1769}

According to Pingre, on September 4 1769 he clearly saw the ripples in the tail similar to that seen in the aurora borealis. The tail was curved, its convex side faced toward the North. Sometimes the tail formed in its extremity a smaller arc, the convexity of which faced to the South, as translated by us from \textcolor{orange}{Pingre (\citeyear{1784coth.book.....P}}, ``des ondulations dans la queue, analogues \`{a} celles que l'on apercoit dans les aurores bor\'{e}ales... La queue \`{e}toit fenfibiement arqu\`{e}e, sa partie convexe tourn\`{e}e vers le nord; quelquefois m\^{e}me elle formoit vers fon extr\`{e}mit\`{e} un second arc plus petit, dont la convexit\`{e} regardoit le sud...''). Pingre also referred to the observations by La Nux who reported 97$^{\circ}$ length of the tail on September 11. Curving cometary tail was also reported by \textcolor{orange}{\citet{Cassini1773}}.

According to Messier, on the night from 4 to 5 September, the sky was serene. Between the first and second hours of the morning, the comet was clearly visible. The tail was 43$^{\circ}$ long, located below the third Orion star. The distant half of the tail shone with very weak light. Also, the tail was curved with bulge to the North. Up to 8$^{\circ}$ from the nucleus, the tail was intermittent along its length by luminous and dim parts. The nucleus had the same color as the day before, it was white with slightly reddish or orange tint \textcolor{orange}{Messier (\citeyear{Messier1778}}, ``La nuit du 4 au 5 de Septembre, le ciel serein; entre une heure \& deux heures du matin, on apercevoit tr\`{e}s-distictement la Com\`{e}te \`{a} la vue simple, avec une queue tr\`{e}s-longue, qui passoit au-dessous de la troisi\`{e}me \'{e}toile d'Orion... ayant quarante-trois degr\'{e}s; plus de la moiti\'{e} de la queue vers sa fin \'{e}toit d'une lumi\`{e}re tr\`{e}s-foible; ele se courboit sensiblement, \& la convexit\'{e} \'{e}toit tourn\'{e}e vers le Nord. La queue depuis le noyau jusqu'\`{a} huit degr\'{e}s de distance environ, \`{e}toit partag\'{e}e suivant sa longueur par des parties lumineuses, \& par d'autres qui \'{e}toient obscures; ces traces lumineuses \& obscures \'{e}toient dans des directions parall\`{e}les. Le noyau de la Com\`{e}te \'{e}toit de la m\^{e}me couleur que la nuit pr\'{e}c\'{e}dente, d'une lumi\`{e}re blanch\^{a}tre, tirant un peu sur le rouge ou l'orang\'{e}...'').

Figure~\textcolor{orange}{\ref{Fig6}b} shows the fragment of the drawing in a concave cylindrical projection by \textcolor{orange}{\citet{Messier1778}}. On contrary to the text description, the comet was depicted with straight tail. On August 30 and September 2 1769, the cometary tail had complex structure. On the right and left of the main tail, Messier painted two short tails less than 5$^{\circ}$ in length (red blocks in Figure~\textcolor{orange}{\ref{Fig6}b}).

For the bright and long main tail, \textcolor{orange}{\citet{Bredichin1880,Bredichin1886}} defined $R=12$ and $\epsilon$ varied from $-5^{\circ}$ to 13$^{\circ}$, on average $\epsilon \approx 4^{\circ}$. He discussed the type of the tail (I or II). Approaching the perihelion (October 8 1769), the nucleolus became bright, but the tail became dim. Therefore, the calculations were made from August 20 to September 12 and from October 25 to November 27 1769.

The small average value of $\epsilon$ signifies that observers might have seen plasma tail whose ray structure indicates non-stationary flow of the solar wind. Large scatter of $\epsilon$ might indicate either sudden changes in the solar wind speed, or uncertainty in observations and further calculations. To clarify this issue, more sophisticated research is required.

Messier also reported more than 10 comets in the second half of the eighteenth century. We analysed their descriptions and drawings. Unfortunately, cometary tails were too short to defined their nature.

\subsection{Catalogues of Hevelius and Lubieniecki} \label{sec:HevLuv}

The catalogues of Hevelius and Lubieniecki collect comets that have been observed from ancient times. Cometographia \textcolor{orange}{\citep{1668ctnc.book.....H}} is devoted to analysis of the variety of cometary nuclei and tails in order to find regularities and explain them. He considered the structure of the sunspots observed by \textcolor{orange}{\citet{scheiner1630}} to reveal the common structural features and matter from which they consist with those of comets. (\textcolor{orange}{Hevelius, \citeyear{1668ctnc.book.....H}}, Figure~G therein) shows the author's poetic cometary description in the distant past. The first object in Figure~G of Cometographia is the artistic imagination of a comet (``Cometan Solarem, seu Rosam'') portrayed as a flame- or red-coloured rose (``colore igneo seu rutilo''); the second object is a comet with bright flame rays. Comets were also compared to shields (``clypei'', the forth object in Figure~G), hair, or mane movements (``Crines'', the seventh object), lamps or facula with straight rays (``adinstar Lampadis, seu facul{\ae} ardentis'', objects 11\,--\,13). Figure~H~and~I depict comets with tails. All these drawings are the author's vision of the shape of comets, according to ancient observations since the fifth century. In particular, object number 28 is the comet which was seen in 1099 and described as a scimitar or curved Persian sword. Cometary nucleus is depicted like a hilt of a sword. Objects 37 and 38 in Figure~K portrayed vibrating ``vibrantibus'') tails. Hevelius also reported comets of the seventeenth century and his own observations. He discussed the parabolic, hyperbolic, and other forms of asymptotes of cometary tails (\textcolor{orange}{Hevelius, \citeyear{1668ctnc.book.....H}}, Figure~Q therein).

We also analyse Historia Cometarum \textcolor{orange}{\citep{Lubieniecii1666}} and Teatrum Comticum by \textcolor{orange}{\citep{Lubienietski1681}}. The author focused on the position of comets in the starry sky. Cometary tails are schematically shown as wide and straight. 

Here we would like to conclude that the mentioned catalogues are not fruitful material for discussion of the plasma tails.

\section{Conclusions} \label{S-Conclusions}

Plasma tails of comets are known as natural probes of the interplanetary space. Their physical parameters may shed light on the solar wind speed in the most intriguing period, the Maunder minimum.

In this paper, we considered descriptions and drawings of comets in the historical archives. To define whether plasma tails were reported in the distant past, we addressed to color, shape, and orientation of cometary tails. 

Reviewing color of comets in European, Chinese (in the translations of Williams and Kronk), and Russian observations, we figured out a lack of convincing facts that naked eye observers saw bluish plasma tails. This finding is trivial, because bluish color originates from the violet bands of ionized carbon monoxide and is poorly recognized by the human eye.

Another crucial attribute of plasma tails is their antisolar orientation. The angle $\epsilon$ between the observed tail and the prolonged radius-vector of a comet is typically limited to 6$^{\circ}$. Here, we provided Bredikhin's estimations of $\epsilon$. Average deviation of cometary tails from the antisolar vector exceeds 10$^{\circ}$, that in turn suggests to us that observers reported dust tails.

Differentiation of dust and plasma tails by their shape should be performed taking into account the projection effects. When a comet passes near the ecliptic plane, the dust tail appears straight and thin. Short dust tails also seem almost straight.

One may conjecture that solar cycle minima represent conditions similar to those during the MM, whose activity level became hotly debating again  \textcolor{orange}{\citep{2015A&A...581A..95U,2016SoPh..291.2869Z}}. A prolonged minimum of Cycles 23\,--\,24 was remarkably spotless, but the Sun was not entirely shrouded in a coronal hole, and the solar wind speed was although suppressed but not crucially. Moreover, \textcolor{orange}{\citet{2007ApJ...657.1137L}} showed that the basal quiet atmosphere is unaffected by solar cycle. If so, we would like to hypothesize that comets in the distant past exhibited plasma tails, but they were not reported because they are hardly seen by the naked eye. Accumulating the light, telescopes could help a little to make out dark-dark bluish plasma tail in the night sky. However, early telescopes likely suffered from spherical and chromatic aberration \textcolor{orange}{\citep{2017SoPh..292....4S}}. Therefore, the first hint of plasma tail we figured out only for Great comet C/1769~P1.



\bibliographystyle{spr-mp-sola}
\bibliography{Zolotova_et_al}  

\begin{thebibliography}{82}
\ifx\bisbn     \undefined \def\bisbn  #1{ISBN #1}\fi
\ifx\binits    \undefined \def\binits#1{#1}\fi
\ifx\bauthor   \undefined \def\bauthor#1{#1}\fi
\ifx\batitle   \undefined \def\batitle#1{#1}\fi
\ifx\bjtitle   \undefined \def\bjtitle#1{\textit{#1}}\fi
\ifx\bvolume   \undefined \def\bvolume#1{\textbf{#1}}\fi
\ifx\byear     \undefined \def\byear#1{#1}\fi
\ifx\bissue    \undefined \def\bissue#1{#1}\fi
\ifx\bfpage    \undefined \def\bfpage#1{#1}\fi
\ifx\blpage    \undefined \def\blpage #1{#1}\fi
\ifx\burl      \undefined \def\burl#1{\textsf{#1}}\fi
\ifx\href      \undefined \def\href#1#2{\textsf{#2}}\fi
\ifx\betal     \undefined \def\betal{\textit{et al.}}\fi
\ifx\bctitle   \undefined \def\bctitle#1{#1}\fi
\ifx\beditor   \undefined \def\beditor#1{#1}\fi
\ifx\bbtitle   \undefined \def\bbtitle#1{\textit{#1}}\fi
\ifx\bedition  \undefined \def\bedition#1{#1}\fi
\ifx\bseriesno \undefined \def\bseriesno#1{\textbf{#1}}\fi
\ifx\blocation \undefined \def\blocation#1{#1}\fi
\ifx\bsertitle \undefined \def\bsertitle#1{\textit{#1}}\fi
\ifx\bsnm      \undefined \def\bsnm#1{#1}\fi
\ifx\bsuffix   \undefined \def\bsuffix#1{#1}\fi
\ifx\bparticle \undefined \def\bparticle#1{#1}\fi
\ifx\barticle  \undefined \def\barticle#1{}\fi
\ifx\binstitute  \undefined \def\binstitute#1{#1}\fi
\ifx\bpublisher  \undefined \def\bpublisher#1{#1}\fi
\ifx\doiurl    \undefined
  \def\doiurl#1{\href{http://dx.doi.org/#1}{\textsf{DOI}}}\fi
\ifx\arxivurl  \undefined
  \def\arxivurl#1{\href{http://arxiv.org/abs/#1}{\textsf{arXiv}}}\fi
\ifx\adsurl    \undefined
  \def\adsurl#1{\href{http://adsabs.harvard.edu/abs/#1}{\textsf{ADS}}}\fi
\ifx\botherref \undefined \def\botherref#1{}\fi
\ifx\url       \undefined \def\url#1{\textsf{#1}}\fi
\ifx\bchapter  \undefined \def\bchapter#1{}\fi
\ifx\bbook     \undefined \def\bbook#1{}\fi
\ifx\bcomment  \undefined \def\bcomment#1{#1}\fi
\ifx\oauthor   \undefined \def\oauthor#1{#1}\fi
\ifx\citeauthoryear \undefined\def \citeauthoryear#1{#1}\fi
\ifx\endbibitem\undefined \def\endbibitem{}\fi
\ifx\bconflocation  \undefined \def\bconflocation#1{#1} \fi

\bibitem[\protect\citeauthoryear{{Bainbridge}}{1619}]{Bainbridge1619}
\begin{bbook}
\bauthor{\bsnm{{Bainbridge}}, \binits{J.}}:
\byear{1619},
\bbtitle{{An Astronomicall description of the late Comet from the 18. of
  Nouemb. 1618 to the 16. of December following}},
\bpublisher{Printed by Edward Griffin},
\blocation{London}.
\end{bbook}
\endbibitem

\bibitem[\protect\citeauthoryear{{Bazelio}}{1578}]{Bazelio1578}
\begin{bbook}
\bauthor{\bsnm{{Bazelio}}, \binits{N.}}:
\byear{1578},
\bbtitle{Prognosticon novum, anni huius calamitosissimi 1578},
\bpublisher{Apud Henricum Henricium},
\blocation{Antverpi{\ae}}.
\end{bbook}
\endbibitem

\bibitem[\protect\citeauthoryear{{Bessel}}{1836}]{1836AnP...114..498B}
\begin{barticle}
\bauthor{\bsnm{{Bessel}}, \binits{F.W.}}:
\byear{1836},
\batitle{{Beobachtungen {\"u}ber die physische Beschaffenheit des Halley'schen
  Kometen und dadurch veranlasste Bemerkungen}}.
\bjtitle{Annalen der Physik}
\bvolume{114},
\bfpage{498}.
\doiurl{10.1002/andp.18361140704}.
\adsurl{1836AnP...114..498B}.
\end{barticle}
\endbibitem

\bibitem[\protect\citeauthoryear{{Biermann}}{1951}]{1951ZA.....29..274B}
\begin{barticle}
\bauthor{\bsnm{{Biermann}}, \binits{L.}}:
\byear{1951},
\batitle{{Kometenschweife und solare Korpuskularstrahlung}}.
\bjtitle{Zeitschrift f\"{u}r Astrophysik}
\bvolume{29},
\bfpage{274}.
\adsurl{1951ZA.....29..274B}.
\end{barticle}
\endbibitem

\bibitem[\protect\citeauthoryear{{Brahe}}{1610}]{brahe1610}
\begin{bbook}
\bauthor{\bsnm{{Brahe}}, \binits{T.}}:
\byear{1610},
\bbtitle{Tychonis brahe dani de mundi {\ae}therei recentioribus ph{\ae}nomenis.
  liber secundus},
\bpublisher{Godefridum Tampachium},
\blocation{Francofurti}.
\end{bbook}
\endbibitem

\bibitem[\protect\citeauthoryear{{Brandt}}{1967}]{1967ApJ...147..201B}
\begin{barticle}
\bauthor{\bsnm{{Brandt}}, \binits{J.C.}}:
\byear{1967},
\batitle{{Interplanetary Gas. XIII. Gross Plasma Velocities from the
  Orientations of Ionic Comet Tails}}.
\bjtitle{\apj}
\bvolume{147},
\bfpage{201}.
\doiurl{10.1086/148992}.
\adsurl{1967ApJ...147..201B}.
\end{barticle}
\endbibitem

\bibitem[\protect\citeauthoryear{{Bredikhin}}{1862}]{Bredikhin1862}
\begin{bbook}
\bauthor{\bsnm{{Bredikhin}}, \binits{F.A.}}:
\byear{1862},
\bbtitle{{On the tails of comets}},
\bpublisher{University printing house},
\blocation{Moscow}.
\end{bbook}
\endbibitem

\bibitem[\protect\citeauthoryear{{Bredikhin}}{1879a}]{Bredichin1879a}
\begin{bbook}
\bauthor{\bsnm{{Bredikhin}}, \binits{T.}}:
\byear{1879}a,
\bbtitle{{Annales de l'observatoire de Moscou, vol. 5, livr. 2}},
\bpublisher{imprimerie F.~Neub\"{u}rger},
\blocation{Moscou}.
\end{bbook}
\endbibitem

\bibitem[\protect\citeauthoryear{{Bredikhin}}{1879b}]{Bredichin1879b}
\begin{bbook}
\bauthor{\bsnm{{Bredikhin}}, \binits{T.}}:
\byear{1879}b,
\bbtitle{{Annales de l'observatoire de Moscou, vol. 6, livr. 1}},
\bpublisher{imprimerie F.~Neub\"{u}rger},
\blocation{Moscou}.
\end{bbook}
\endbibitem

\bibitem[\protect\citeauthoryear{{Bredikhin}}{1880}]{Bredichin1880}
\begin{bbook}
\bauthor{\bsnm{{Bredikhin}}, \binits{T.}}:
\byear{1880},
\bbtitle{{Annales de l'observatoire de Moscou, vol. 7, livr. 1}},
\bpublisher{imprimerie F.~Neub\"{u}rger},
\blocation{Moscou}.
\end{bbook}
\endbibitem

\bibitem[\protect\citeauthoryear{{Bredikhin}}{1886}]{Bredichin1886}
\begin{bbook}
\bauthor{\bsnm{{Bredikhin}}, \binits{T.}}:
\byear{1886},
\bbtitle{{Annales de l'observatoire de Moscou, deuxieme serie, vol. 1, livr.
  1}},
\bpublisher{imprimerie F~ Neub\"{u}rger},
\blocation{Moscou}.
\end{bbook}
\endbibitem

\bibitem[\protect\citeauthoryear{{Busch}}{1577}]{Busch1577}
\begin{bbook}
\bauthor{\bsnm{{Busch}}, \binits{G.}}:
\byear{1577},
\bbtitle{Beschreibung / von zugeh\"{o}rigen eigenschafften / und natürlicher
  jnfluentz / des grossen und erschrecklichen cometen / welcher in diesem
  1577},
\bpublisher{Mechler},
\blocation{Nuremberg}.
\end{bbook}
\endbibitem

\bibitem[\protect\citeauthoryear{{Cassini}}{1665}]{Cassini1665}
\begin{bbook}
\bauthor{\bsnm{{Cassini}}, \binits{D.J.}}:
\byear{1665},
\bbtitle{Theoriae motus cometae anni mdclxiv..},
\bpublisher{Ex typographia Fabii de Falco},
\blocation{Romae}.
\end{bbook}
\endbibitem

\bibitem[\protect\citeauthoryear{{Cassini}}{1681}]{Cassini1681}
\begin{bbook}
\bauthor{\bsnm{{Cassini}}, \binits{G.D.}}:
\byear{1681},
\bbtitle{Abreg\'{e} des observations \& des reflexions sur la comete qui a paru
  au mois de decembre 1680..},
\bpublisher{Michallet},
\blocation{Paris}.
\end{bbook}
\endbibitem

\bibitem[\protect\citeauthoryear{{Cassini}}{1730}]{Cassini1730}
\begin{barticle}
\bauthor{\bsnm{{Cassini}}, \binits{G.D.}}:
\byear{1730},
\batitle{{Nouve au Phenomene rare et singulier...}}
\bjtitle{M\'{e}moires de L'Acad\'{e}mie Royale des Science. Depuis 1666 jusqu'
  \`{a} 1699}
\bvolume{{Tome X}},
\bfpage{637}.
\end{barticle}
\endbibitem

\bibitem[\protect\citeauthoryear{{Cassini de Thury}}{1773}]{Cassini1773}
\begin{barticle}
\bauthor{\bsnm{{Cassini de Thury}}, \binits{C.F.}}:
\byear{1773},
\batitle{{Observation et th\'{e}orie de la Com\`{e}te qui a paru au mois
  d'Ao\^{u}t 1769, avec quelques R\'{e}flexions sur les th\'{e}ories d'une
  même Com\`{e}te, \'{e}tablies dans diff\'{e}rentes apparitions}}.
\bjtitle{Histoire de L'Acad\'{e}mie Royale des Science}
\bvolume{1770},
\bfpage{24}.
\end{barticle}
\endbibitem

\bibitem[\protect\citeauthoryear{{Cheseaux}}{1744}]{Cheseaux1744}
\begin{bbook}
\bauthor{\bsnm{{Cheseaux}}, \binits{J.P.L.}}:
\byear{1744},
\bbtitle{Trait\'{e} de la comete qui a paru en decembre 1743},
\bpublisher{Chez Marc-Michel Bousquet \& compagnie},
\blocation{A Lausanne \& \`{a} Geneve}.
\end{bbook}
\endbibitem

\bibitem[\protect\citeauthoryear{{Cliver}}{2012}]{2012IAUS..286..179C}
\begin{bchapter}
\bauthor{\bsnm{{Cliver}}, \binits{E.W.}}:
\byear{2012},
\bctitle{{The floor in the solar wind: status report}}.
In: \beditor{\bsnm{{Mandrini}}, \binits{C.H.}},
\beditor{\bsnm{{Webb}}, \binits{D.F.}} (eds.)
\bbtitle{Comparative Magnetic Minima: Characterizing Quiet Times in the Sun and
  Stars},
\bsertitle{IAU Symposium}
\bseriesno{286},
\bfpage{179}.
\doiurl{10.1017/S1743921312004814}.
\adsurl{2012IAUS..286..179C}.
\end{bchapter}
\endbibitem

\bibitem[\protect\citeauthoryear{{Cliver}, {Boriakoff}, and
  {Bounar}}{1998}]{1998GeoRL..25..897C}
\begin{barticle}
\bauthor{\bsnm{{Cliver}}, \binits{E.W.}},
\bauthor{\bsnm{{Boriakoff}}, \binits{V.}},
\bauthor{\bsnm{{Bounar}}, \binits{K.H.}}:
\byear{1998},
\batitle{{Geomagnetic activity and the solar wind during the Maunder Minimum}}.
\bjtitle{\grl}
\bvolume{25},
\bfpage{897}.
\doiurl{10.1029/98GL00500}.
\adsurl{1998GeoRL..25..897C}.
\end{barticle}
\endbibitem

\bibitem[\protect\citeauthoryear{{Cysato}}{1619}]{Cysato1619}
\begin{bbook}
\bauthor{\bsnm{{Cysato}}, \binits{I.B.}}:
\byear{1619},
\bbtitle{Mathematica astronomica de loco, motu, magnitudine et causis cometae
  qui sub finem anni 1619},
\bpublisher{Ex Typographeo Ederiano},
\blocation{Ingolstadii}.
\end{bbook}
\endbibitem

\bibitem[\protect\citeauthoryear{{Eddy}}{1976}]{1976Sci...192.1189E}
\begin{barticle}
\bauthor{\bsnm{{Eddy}}, \binits{J.A.}}:
\byear{1976},
\batitle{{The Maunder Minimum}}.
\bjtitle{Science}
\bvolume{192},
\bfpage{1189}.
\doiurl{10.1126/science.192.4245.1189}.
\adsurl{1976Sci...192.1189E}.
\end{barticle}
\endbibitem

\bibitem[\protect\citeauthoryear{{Euler}}{1744}]{Euler1744}
\begin{bbook}
\bauthor{\bsnm{{Euler}}, \binits{L.}}:
\byear{1744},
\bbtitle{Fortgesetzte beantwortung der fragen \"{u}ber die beschaffenheit,
  bewegung und w\"{u}rkung der cometen},
\bpublisher{Haude},
\blocation{Berlin}.
\end{bbook}
\endbibitem

\bibitem[\protect\citeauthoryear{{Flamsteed}}{1725}]{Flamsteed1725}
\begin{bbook}
\bauthor{\bsnm{{Flamsteed}}, \binits{J.}}:
\byear{1725},
\bbtitle{Histori{\ae} coelestis britannic{\ae}. volumen primum.},
\bpublisher{Typis H. Meere},
\blocation{Londini}.
\end{bbook}
\endbibitem

\bibitem[\protect\citeauthoryear{{Graminaeus}}{1578}]{Graminaeum1578}
\begin{bbook}
\bauthor{\bsnm{{Graminaeus}}, \binits{T.}}:
\byear{1578},
\bbtitle{Weltspiegel oder/ algemeiner widerwertigkeit/ dess f\"{u}nfften
  kirchen alters/ k\"{u}rtze verzeignuss},
\bpublisher{Durch Ludouicum Alectorium},
\blocation{C\"{o}lln}.
\end{bbook}
\endbibitem

\bibitem[\protect\citeauthoryear{{Graminaeus}}{1581}]{Graminaeus1581}
\begin{bbook}
\bauthor{\bsnm{{Graminaeus}}, \binits{T.}}:
\byear{1581},
\bbtitle{Comet{\ae} anni dimini 1580},
\bpublisher{Administratoris Episcopatus Monasteriensis},
\blocation{Coloni{\ae} Agrippin{\ae}}.
\end{bbook}
\endbibitem

\bibitem[\protect\citeauthoryear{{Gulyaev}}{2015}]{2015ARep...59..791G}
\begin{barticle}
\bauthor{\bsnm{{Gulyaev}}, \binits{R.A.}}:
\byear{2015},
\batitle{{Type I cometary tails and the solar wind at the epoch of the Maunder
  minimum}}.
\bjtitle{Astronomy Reports}
\bvolume{59},
\bfpage{791}.
\doiurl{10.1134/S106377291508003X}.
\adsurl{2015ARep...59..791G}.
\end{barticle}
\endbibitem

\bibitem[\protect\citeauthoryear{{Hayck}}{1578}]{Hayck1578}
\begin{bbook}
\bauthor{\bsnm{{Hayck}}, \binits{T.H.}}:
\byear{1578},
\bbtitle{{Descriptio comet{\ae}, qui apparuit anno domini M.D.LXXVII. \`{a} IX.
  die Novembris usque ad XIII. diem Ianuarij, Anni \&c. LXXVIII}},
\bpublisher{Georgii Melantrichi},
\blocation{Prag{\ae}}.
\end{bbook}
\endbibitem

\bibitem[\protect\citeauthoryear{{Heinsius}}{1744}]{Heinsius1744}
\begin{bbook}
\bauthor{\bsnm{{Heinsius}}, \binits{G.}}:
\byear{1744},
\bbtitle{Beschreibung des im anfang des jahrs 1744 erschienenen cometen...},
\bpublisher{Academie derer Wissenschafften},
\blocation{St. Petersburg}.
\end{bbook}
\endbibitem

\bibitem[\protect\citeauthoryear{{Hevelius}}{1665}]{Hevelii1665}
\begin{bbook}
\bauthor{\bsnm{{Hevelius}}, \binits{J.}}:
\byear{1665},
\bbtitle{Prodromus cometicus quo historia comet{\ae} anno 1664..},
\bpublisher{Autoris Typis, Et Sumptibus},
\blocation{Gedani}.
\end{bbook}
\endbibitem

\bibitem[\protect\citeauthoryear{{Hevelius}}{1666}]{Hevelii1666}
\begin{bbook}
\bauthor{\bsnm{{Hevelius}}, \binits{J.}}:
\byear{1666},
\bbtitle{Descriptio comet{\ae} anno {\ae}rae christ. m.dc.lxv},
\bpublisher{Autoris Typis, Et Sumptibus},
\blocation{Gedani}.
\end{bbook}
\endbibitem

\bibitem[\protect\citeauthoryear{{Hevelius}}{1668}]{1668ctnc.book.....H}
\begin{bbook}
\bauthor{\bsnm{{Hevelius}}, \binits{J.}}:
\byear{1668},
\bbtitle{{Cometographia, totam naturam cometarum ...}},
\bpublisher{imprimebat S.~Reiniger},
\blocation{Gedani}.
\adsurl{1668ctnc.book.....H}.
\end{bbook}
\endbibitem

\bibitem[\protect\citeauthoryear{{Hevelius}}{1672}]{Hevelius1672}
\begin{bbook}
\bauthor{\bsnm{{Hevelius}}, \binits{J.}}:
\byear{1672},
\bbtitle{{Epistola de cometa, anno M DC LXXII, Mense Martio, \& Aprili, Gedani
  Observato}},
\bpublisher{Autoris Typis, \& sumptibus Imprimebat, Simon Reoniger},
\blocation{Gedani}.
\end{bbook}
\endbibitem

\bibitem[\protect\citeauthoryear{{Hevelius}}{1685}]{Hevelii1685}
\begin{bbook}
\bauthor{\bsnm{{Hevelius}}, \binits{J.}}:
\byear{1685},
\bbtitle{Annus climactericus},
\bpublisher{Sumptibus Auctoris, Typis Dav. Frid. Rhetii},
\blocation{Gedani}.
\end{bbook}
\endbibitem

\bibitem[\protect\citeauthoryear{{Honold}}{1682}]{Honold1682}
\begin{bbook}
\bauthor{\bsnm{{Honold}}, \binits{G.}}:
\byear{1682},
\bbtitle{Kurzer entwurf des neuentstandenen cometen... 1682},
\bpublisher{Erben},
\blocation{Ulm}.
\end{bbook}
\endbibitem

\bibitem[\protect\citeauthoryear{{Honold}}{1677}]{Honold1677}
\begin{bbook}
\bauthor{\bsnm{{Honold}}, \binits{J.}}:
\byear{1677},
\bbtitle{Sidereus dei clarigator, das ist kurtzer bericht vom dem neuen cometen
  welcher in der oesterlichen zeit dieses 1677},
\bpublisher{Christian Baltbasar Bussnen},
\blocation{Ulm}.
\end{bbook}
\endbibitem

\bibitem[\protect\citeauthoryear{{Hooke}}{1678}]{Hooke1678}
\begin{bbook}
\bauthor{\bsnm{{Hooke}}, \binits{R.}}:
\byear{1678},
\bbtitle{Lectures and collections... Сometa},
\bpublisher{John Martyn},
\blocation{London}.
\end{bbook}
\endbibitem

\bibitem[\protect\citeauthoryear{{Kepler}}{1619}]{Keplero1619}
\begin{bbook}
\bauthor{\bsnm{{Kepler}}, \binits{I.}}:
\byear{1619},
\bbtitle{De cometis libelli tres},
\bpublisher{typis Andrae Apergeri},
\blocation{August{\ae} Vindelicorum}.
\end{bbook}
\endbibitem

\bibitem[\protect\citeauthoryear{{Kirch}}{1682}]{Kirch1682}
\begin{bbook}
\bauthor{\bsnm{{Kirch}}, \binits{G.}}:
\byear{1682},
\bbtitle{Eilfertiger kurtzer bericht an einen guten freund von dem neuen
  cometen dieses 1682},
\bpublisher{Endter},
\blocation{N\"{u}rnberg}.
\end{bbook}
\endbibitem

\bibitem[\protect\citeauthoryear{{Kronk}}{1984}]{1984cdc..book.....K}
\begin{bbook}
\bauthor{\bsnm{{Kronk}}, \binits{G.W.}}:
\byear{1984},
\bbtitle{{Comets: a descriptive catalog}},
\bpublisher{Enslow Publishers},
\blocation{Hillside, N.J., U.S.A.}.
\end{bbook}
\endbibitem

\bibitem[\protect\citeauthoryear{{Kronk}}{1999}]{1999ccc..book.....K}
\begin{bbook}
\bauthor{\bsnm{{Kronk}}, \binits{G.W.}}:
\byear{1999},
\bbtitle{{Cometography: A Catalog of Comets, Volume 1: Ancient-1799}},
\bpublisher{Cambridge University Press},
\blocation{Cambridge}.
\adsurl{1999ccc..book.....K}.
\end{bbook}
\endbibitem

\bibitem[\protect\citeauthoryear{{Livingston}
  \textit{et~al.}}{2007}]{2007ApJ...657.1137L}
\begin{barticle}
\bauthor{\bsnm{{Livingston}}, \binits{W.}},
\bauthor{\bsnm{{Wallace}}, \binits{L.}},
\bauthor{\bsnm{{White}}, \binits{O.R.}},
\bauthor{\bsnm{{Giampapa}}, \binits{M.S.}}:
\byear{2007},
\batitle{{Sun-as-a-Star Spectrum Variations 1974-2006}}.
\bjtitle{\apj}
\bvolume{657},
\bfpage{1137}.
\doiurl{10.1086/511127}.
\adsurl{2007ApJ...657.1137L}.
\end{barticle}
\endbibitem

\bibitem[\protect\citeauthoryear{{Lockwood} and
  {Owens}}{2014}]{2014ApJ...781L...7L}
\begin{barticle}
\bauthor{\bsnm{{Lockwood}}, \binits{M.}},
\bauthor{\bsnm{{Owens}}, \binits{M.J.}}:
\byear{2014},
\batitle{{Implications of the Recent Low Solar Minimum for the Solar Wind
  during the Maunder Minimum}}.
\bjtitle{\apjl}
\bvolume{781},
\bfpage{L7}.
\doiurl{10.1088/2041-8205/781/1/L7}.
\adsurl{2014ApJ...781L...7L}.
\end{barticle}
\endbibitem

\bibitem[\protect\citeauthoryear{{Lubieniecki}}{1666}]{Lubieniecii1666}
\begin{bbook}
\bauthor{\bsnm{{Lubieniecki}}, \binits{S.}}:
\byear{1666},
\bbtitle{{\`{a} Lubieniec Historia cometarum}},
\bpublisher{Francisco Cupero},
\blocation{prope Portam Harlemensem}.
\end{bbook}
\endbibitem

\bibitem[\protect\citeauthoryear{{Lubieniecki}}{1681}]{Lubienietski1681}
\begin{bbook}
\bauthor{\bsnm{{Lubieniecki}}, \binits{S.}}:
\byear{1681},
\bbtitle{{Theatrum cometicum}},
\bpublisher{Ex officina Petri vander Meersche},
\blocation{Lugduni Batavorum}.
\end{bbook}
\endbibitem

\bibitem[\protect\citeauthoryear{{M{\ae}stlino}}{1578}]{Maestlino1578}
\begin{bbook}
\bauthor{\bsnm{{M{\ae}stlino}}, \binits{M.}}:
\byear{1578},
\bbtitle{Observatio \& demonstratio comet{\ae} {\ae}therei, qui anno 1577 et
  1578},
\bpublisher{Gruppenbachius},
\blocation{Tubing{\ae}}.
\end{bbook}
\endbibitem

\bibitem[\protect\citeauthoryear{{Maraldi}}{1743}]{Maraldi1743}
\begin{barticle}
\bauthor{\bsnm{{Maraldi}}, \binits{G.F.}}:
\byear{1743},
\batitle{{Observation D'un nouveau Ph\'{e}nom\`{e}ne, faite le 2 de Mars 1702,
  par M. Maraldi \`{a} Rome}}.
\bjtitle{Histoire de L'Acad\'{e}mie Royale des Science}
\bvolume{1702},
\bfpage{101}.
\end{barticle}
\endbibitem

\bibitem[\protect\citeauthoryear{{Mayern}}{1681}]{Mayern1681}
\begin{bbook}
\bauthor{\bsnm{{Mayern}}, \binits{J.}}:
\byear{1681},
\bbtitle{Vorstellung dess j\"{u}ngst-erschienenen cometen, wie derselbe vom 16.
  novembris, anno 1680. biss auf den 7. februarii, anno 1681},
\bpublisher{K\"{u}hnen},
\blocation{Dusslingen}.
\end{bbook}
\endbibitem

\bibitem[\protect\citeauthoryear{{Megerlino}}{1661}]{Megerlino1661}
\begin{bbook}
\bauthor{\bsnm{{Megerlino}}, \binits{P.}}:
\byear{1661},
\bbtitle{Discursus mathematicus de cometa nuper viso, in inclyta academia
  basiliensi public\`{e} habitus, die 15. febr. anno 1661},
\bpublisher{Apud Joannem K\"{o}nig},
\blocation{Basileae}.
\end{bbook}
\endbibitem

\bibitem[\protect\citeauthoryear{{Mendis}}{2007}]{2007hste.book..494M}
\begin{bbook}
\bauthor{\bsnm{{Mendis}}, \binits{D.A.}}:
\byear{2007},
In: \beditor{\bsnm{{Kamide}}, \binits{Y.}},
\beditor{\bsnm{{Chian}}, \binits{A.C.-L.}} (eds.)
\bbtitle{{The Solar-Comet Interactions}},
\bfpage{494}.
\adsurl{2007hste.book..494M}.
\end{bbook}
\endbibitem

\bibitem[\protect\citeauthoryear{{Mendoza}}{1997}]{1997AnGeo..15..397M}
\begin{barticle}
\bauthor{\bsnm{{Mendoza}}, \binits{B.}}:
\byear{1997},
\batitle{{Geomagnetic activity and wind velocity during the Maunder minimum}}.
\bjtitle{Annales Geophysicae}
\bvolume{15},
\bfpage{397}.
\doiurl{10.1007/s00585-997-0397-3}.
\adsurl{1997AnGeo..15..397M}.
\end{barticle}
\endbibitem

\bibitem[\protect\citeauthoryear{{Merian}}{1691}]{Merian1691}
\begin{bbook}
\bauthor{\bsnm{{Merian}}, \binits{M.}}:
\byear{1691},
\bbtitle{Theatri europ{\ae}i},
\bpublisher{Matth\"{a}i Merians Sel. Erben},
\blocation{Franchfurt am Mayn}.
\end{bbook}
\endbibitem

\bibitem[\protect\citeauthoryear{{Messier}}{1778}]{Messier1778}
\begin{barticle}
\bauthor{\bsnm{{Messier}}, \binits{M.}}:
\byear{1778},
\batitle{{M\'{e}moire contenant les Observations de la X.e Com\`{e}te
  observ\'{e}e \`{a} Paris, de l'Observatoire de la Marine; depuis le mois
  d'Ao\^{u}t jusqu'au I.er Décembre 1769}}.
\bjtitle{Histoire de L'Acad\'{e}mie Royale des Science}
\bvolume{1775},
\bfpage{392}.
\end{barticle}
\endbibitem

\bibitem[\protect\citeauthoryear{{Montanari}}{1665}]{Montanari1665}
\begin{bbook}
\bauthor{\bsnm{{Montanari}}, \binits{G.}}:
\byear{1665},
\bbtitle{Cometes bononi{\ae} observatus anno 1664 et 1665},
\bpublisher{Typis Io. Baptist{\ae}Ferronij},
\blocation{Bologni{\ae}}.
\end{bbook}
\endbibitem

\bibitem[\protect\citeauthoryear{{Montanari}}{1682}]{Montanari1682}
\begin{bbook}
\bauthor{\bsnm{{Montanari}}, \binits{G.}}:
\byear{1682},
\bbtitle{Copia di lettera scritta all’illustrissimo signore antonio
  magliabechi bibliotecario del serenissimo gran duca di toscana intorno alla
  nuova cometa apparsa quest’ anno 1682},
\bpublisher{Per li Manolessi, Stampatori Camerali},
\blocation{Bologna}.
\end{bbook}
\endbibitem

\bibitem[\protect\citeauthoryear{{Montl'hery}}{1619}]{Montlhery1619}
\begin{bbook}
\bauthor{\bsnm{{Montl'hery}}, \binits{E.}}:
\byear{1619},
\bbtitle{De la trompette du ciel: C' est \`{a} dire, du comete effrayable, qui
  l' an de christ 1618},
\bpublisher{Imprimerie de la Societ\'{e} heluetiale Caldoresque},
\blocation{Iverdon}.
\end{bbook}
\endbibitem

\bibitem[\protect\citeauthoryear{{Orlov}}{1935}]{Orlov1935}
\begin{bbook}
\bauthor{\bsnm{{Orlov}}, \binits{S.V.}}:
\byear{1935},
\bbtitle{{Komety (Comets)}},
\bpublisher{INTI},
\blocation{Moscow}.
\end{bbook}
\endbibitem

\bibitem[\protect\citeauthoryear{{Owens}, {Lockwood}, and
  {Riley}}{2017}]{2017NatSR...741548O}
\begin{barticle}
\bauthor{\bsnm{{Owens}}, \binits{M.J.}},
\bauthor{\bsnm{{Lockwood}}, \binits{M.}},
\bauthor{\bsnm{{Riley}}, \binits{P.}}:
\byear{2017},
\batitle{{Global solar wind variations over the last four centuries}}.
\bjtitle{Scientific Reports}
\bvolume{7},
\bfpage{41548}.
\doiurl{10.1038/srep41548}.
\adsurl{2017NatSR...741548O}.
\end{barticle}
\endbibitem

\bibitem[\protect\citeauthoryear{{Parker}}{1958}]{1958ApJ...128..664P}
\begin{barticle}
\bauthor{\bsnm{{Parker}}, \binits{E.N.}}:
\byear{1958},
\batitle{{Dynamics of the Interplanetary Gas and Magnetic Fields.}}
\bjtitle{\apj}
\bvolume{128},
\bfpage{664}.
\doiurl{10.1086/146579}.
\adsurl{1958ApJ...128..664P}.
\end{barticle}
\endbibitem

\bibitem[\protect\citeauthoryear{{Parker}}{1976}]{1976IAUS...71....3P}
\begin{bchapter}
\bauthor{\bsnm{{Parker}}, \binits{E.N.}}:
\byear{1976},
\bctitle{{The Enigma of Solar Activity}}.
In: \beditor{\bsnm{{Bumba}}, \binits{V.}},
\beditor{\bsnm{{Kleczek}}, \binits{J.}} (eds.)
\bbtitle{Basic Mechanisms of Solar Activity},
\bsertitle{IAU Symposium}
\bseriesno{71},
\bfpage{3}.
\adsurl{1976IAUS...71....3P}.
\end{bchapter}
\endbibitem

\bibitem[\protect\citeauthoryear{{Pingr{\'e}}}{1784}]{1784coth.book.....P}
\begin{bbook}
\bauthor{\bsnm{{Pingr{\'e}}}, \binits{A.G.}}:
\byear{1784},
\bbtitle{{Com{\'e}tographie. Tome second}},
\bpublisher{Imprimerie Royale},
\blocation{Paris}.
\adsurl{1784coth.book.....P}.
\end{bbook}
\endbibitem

\bibitem[\protect\citeauthoryear{{Riley}
  \textit{et~al.}}{2015}]{2015ApJ...802..105R}
\begin{barticle}
\bauthor{\bsnm{{Riley}}, \binits{P.}},
\bauthor{\bsnm{{Lionello}}, \binits{R.}},
\bauthor{\bsnm{{Linker}}, \binits{J.A.}},
\bauthor{\bsnm{{Cliver}}, \binits{E.}},
\bauthor{\bsnm{{Balogh}}, \binits{A.}},
\bauthor{\bsnm{{Beer}}, \binits{J.}},
\bauthor{\bsnm{{Charbonneau}}, \binits{P.}},
\bauthor{\bsnm{{Crooker}}, \binits{N.}},
\bauthor{\bsnm{{DeRosa}}, \binits{M.}},
\bauthor{\bsnm{{Lockwood}}, \binits{M.}},
\bauthor{\bsnm{{Owens}}, \binits{M.}},
\bauthor{\bsnm{{McCracken}}, \binits{K.}},
\bauthor{\bsnm{{Usoskin}}, \binits{I.}},
\bauthor{\bsnm{{Koutchmy}}, \binits{S.}}:
\byear{2015},
\batitle{{Inferring the Structure of the Solar Corona and Inner Heliosphere
  During the Maunder Minimum Using Global Thermodynamic Magnetohydrodynamic
  Simulations}}.
\bjtitle{\apj}
\bvolume{802},
\bfpage{105}.
\doiurl{10.1088/0004-637X/802/2/105}.
\adsurl{2015ApJ...802..105R}.
\end{barticle}
\endbibitem

\bibitem[\protect\citeauthoryear{Scheiner}{1630}]{scheiner1630}
\begin{bbook}
\bauthor{\bsnm{Scheiner}, \binits{C.}}:
\byear{1630},
\bbtitle{{Rosa Ursina}},
\bpublisher{Braccianum},
\blocation{Augustae Vindelicorum}.
\end{bbook}
\endbibitem

\bibitem[\protect\citeauthoryear{{Steinhilber}
  \textit{et~al.}}{2010}]{2010JGRA..115.1104S}
\begin{barticle}
\bauthor{\bsnm{{Steinhilber}}, \binits{F.}},
\bauthor{\bsnm{{Abreu}}, \binits{J.A.}},
\bauthor{\bsnm{{Beer}}, \binits{J.}},
\bauthor{\bsnm{{McCracken}}, \binits{K.G.}}:
\byear{2010},
\batitle{{Interplanetary magnetic field during the past 9300 years inferred
  from cosmogenic radionuclides}}.
\bjtitle{Journal of Geophysical Research (Space Physics)}
\bvolume{115},
\bfpage{A01104}.
\doiurl{10.1029/2009JA014193}.
\adsurl{2010JGRA..115.1104S}.
\end{barticle}
\endbibitem

\bibitem[\protect\citeauthoryear{{Suess}}{1979}]{1979P&SS...27.1001S}
\begin{barticle}
\bauthor{\bsnm{{Suess}}, \binits{S.T.}}:
\byear{1979},
\batitle{{The solar wind during the Maunder minimum}}.
\bjtitle{Planet. Space Sci.}
\bvolume{27},
\bfpage{1001}.
\doiurl{10.1016/0032-0633(79)90030-8}.
\adsurl{1979P\%26SS...27.1001S}.
\end{barticle}
\endbibitem

\bibitem[\protect\citeauthoryear{{Svalgaard}}{2017}]{2017SoPh..292....4S}
\begin{barticle}
\bauthor{\bsnm{{Svalgaard}}, \binits{L.}}:
\byear{2017},
\batitle{{A Recount of Sunspot Groups on Staudach's Drawings}}.
\bjtitle{\solphys}
\bvolume{292},
\bfpage{4}.
\doiurl{10.1007/s11207-016-1023-x}.
\adsurl{2017SoPh..292....4S}.
\end{barticle}
\endbibitem

\bibitem[\protect\citeauthoryear{{Svalgaard} and
  {Cliver}}{2007}]{2007ApJ...661L.203S}
\begin{barticle}
\bauthor{\bsnm{{Svalgaard}}, \binits{L.}},
\bauthor{\bsnm{{Cliver}}, \binits{E.W.}}:
\byear{2007},
\batitle{{A Floor in the Solar Wind Magnetic Field}}.
\bjtitle{\apjl}
\bvolume{661},
\bfpage{L203}.
\doiurl{10.1086/518786}.
\adsurl{2007ApJ...661L.203S}.
\end{barticle}
\endbibitem

\bibitem[\protect\citeauthoryear{Svyatsky}{2007}]{Svyatsky2007}
\begin{bbook}
\bauthor{\bsnm{Svyatsky}, \binits{D.O.}}:
\byear{2007},
\bbtitle{{Astronomy of ancient Russia}},
\bpublisher{Russian pyramid (Russkaya piramida)},
\blocation{Moscow}.
\end{bbook}
\endbibitem

\bibitem[\protect\citeauthoryear{{Usoskin}
  \textit{et~al.}}{2015}]{2015A&A...581A..95U}
\begin{barticle}
\bauthor{\bsnm{{Usoskin}}, \binits{I.G.}},
\bauthor{\bsnm{{Arlt}}, \binits{R.}},
\bauthor{\bsnm{{Asvestari}}, \binits{E.}},
\bauthor{\bsnm{{Hawkins}}, \binits{E.}},
\bauthor{\bsnm{{K{\"a}pyl{\"a}}}, \binits{M.}},
\bauthor{\bsnm{{Kovaltsov}}, \binits{G.A.}},
\bauthor{\bsnm{{Krivova}}, \binits{N.}},
\bauthor{\bsnm{{Lockwood}}, \binits{M.}},
\bauthor{\bsnm{{Mursula}}, \binits{K.}},
\bauthor{\bsnm{{O'Reilly}}, \binits{J.}},
\bauthor{\bsnm{{Owens}}, \binits{M.}},
\bauthor{\bsnm{{Scott}}, \binits{C.J.}},
\bauthor{\bsnm{{Sokoloff}}, \binits{D.D.}},
\bauthor{\bsnm{{Solanki}}, \binits{S.K.}},
\bauthor{\bsnm{{Soon}}, \binits{W.}},
\bauthor{\bsnm{{Vaquero}}, \binits{J.M.}}:
\byear{2015},
\batitle{{The Maunder minimum (1645-1715) was indeed a grand minimum: A
  reassessment of multiple datasets}}.
\bjtitle{\aap}
\bvolume{581},
\bfpage{A95}.
\doiurl{10.1051/0004-6361/201526652}.
\adsurl{2015A\%26A...581A..95U}.
\end{barticle}
\endbibitem

\bibitem[\protect\citeauthoryear{{Uzielli} and {Celoria}}{1893}]{Uzielli1893}
\begin{bbook}
\bauthor{\bsnm{{Uzielli}}, \binits{G.}},
\bauthor{\bsnm{{Celoria}}, \binits{G.}}:
\byear{1893},
\bbtitle{Raccolta di documenti e studi pubblicati dalla r. commissione
  colombiana pel quarto centenario dalla scoperta dell'america. volumen 10.
  parte v. la vita e i tempi di paolo dal pozzo toscanelli},
\bpublisher{Ben Jihann Francisco Bortoletti},
\blocation{Jena}.
\end{bbook}
\endbibitem

\bibitem[\protect\citeauthoryear{{Voigt}}{1677}]{Voigt1677}
\begin{bbook}
\bauthor{\bsnm{{Voigt}}, \binits{J.H.}}:
\byear{1677},
\bbtitle{Christm\"{a}ssige betrachtung des cometen im aprili anno 1677},
\bpublisher{Georg Rebenlein},
\blocation{Hamburg}.
\end{bbook}
\endbibitem

\bibitem[\protect\citeauthoryear{{Voigt}}{1681}]{Voigt1681}
\begin{bbook}
\bauthor{\bsnm{{Voigt}}, \binits{J.H.}}:
\byear{1681},
\bbtitle{Cometa matutinus \& vespertinus... anno 1680. und 1681},
\bpublisher{Rebenlein},
\blocation{Gamburg}.
\end{bbook}
\endbibitem

\bibitem[\protect\citeauthoryear{{Wang} and
  {Sheeley}}{2013}]{2013ApJ...764...90W}
\begin{barticle}
\bauthor{\bsnm{{Wang}}, \binits{Y.-M.}},
\bauthor{\bsnm{{Sheeley}}, \binits{N.R.} \bsuffix{Jr.}}:
\byear{2013},
\batitle{{The Solar Wind and Interplanetary Field during Very Low Amplitude
  Sunspot Cycles}}.
\bjtitle{\apj}
\bvolume{764},
\bfpage{90}.
\doiurl{10.1088/0004-637X/764/1/90}.
\adsurl{2013ApJ...764...90W}.
\end{barticle}
\endbibitem

\bibitem[\protect\citeauthoryear{{Weigel}}{1653}]{Weigel1653}
\begin{bbook}
\bauthor{\bsnm{{Weigel}}, \binits{E.}}:
\byear{1653},
\bbtitle{Commentatio astronomica de cometa novo. qui sub finem anni 1652,
  lumine subobscuro nobis illuxit},
\bpublisher{Typis Georgii Sengenuualdi},
\blocation{Jena}.
\end{bbook}
\endbibitem

\bibitem[\protect\citeauthoryear{{Weigel}}{1672}]{Weigel1672}
\begin{bbook}
\bauthor{\bsnm{{Weigel}}, \binits{E.}}:
\byear{1672},
\bbtitle{Erhardi weigelii vorstellung der kunst und handwerke},
\bpublisher{Johann Jacob Bauhofern},
\blocation{Jena}.
\end{bbook}
\endbibitem

\bibitem[\protect\citeauthoryear{{Welper}}{1661}]{Welper1661}
\begin{bbook}
\bauthor{\bsnm{{Welper}}, \binits{E.}}:
\byear{1661},
\bbtitle{Cometographia. oder beschreibung dessen im 1661},
\bpublisher{durch M. Eberhard Welpern, Mathematicum},
\blocation{Strassburg}.
\end{bbook}
\endbibitem

\bibitem[\protect\citeauthoryear{{Wiedeburg}}{1742}]{Wiedeburg1742}
\begin{bbook}
\bauthor{\bsnm{{Wiedeburg}}, \binits{J.B.}}:
\byear{1742},
\bbtitle{Astronomische beschreibung und nachricht von dem cometen, welcher im
  monath mertz dieses jetztlauffenden jahres 1742},
\bpublisher{Ben Jihann Francisco Bortoletti},
\blocation{Jena}.
\end{bbook}
\endbibitem

\bibitem[\protect\citeauthoryear{{Williams}}{1871}]{1871obco.book.....W}
\begin{bbook}
\bauthor{\bsnm{{Williams}}, \binits{J.}}:
\byear{1871},
\bbtitle{{Observations of comets, from B.C. 611 to A.D. 1640, Extracted from
  the Chinese Annals}},
\bpublisher{Science and Technology Publishers LTD},
\blocation{hornchurch, Essex, England}.
\end{bbook}
\endbibitem

\bibitem[\protect\citeauthoryear{{Wurm}}{1954}]{Wurm1954}
\begin{bbook}
\bauthor{\bsnm{{Wurm}}, \binits{K.}}:
\byear{1954},
\bbtitle{{Die Kometen}},
\bpublisher{Springer},
\blocation{Berlin, Heidelberg}.
\doiurl{https://doi.org/10.1007/978-3-642-86294-6}.
\end{bbook}
\endbibitem

\bibitem[\protect\citeauthoryear{{Zanotti}}{1739}]{Zanotti1739}
\begin{bbook}
\bauthor{\bsnm{{Zanotti}}, \binits{E.}}:
\byear{1739},
\bbtitle{La cometa dell' anno mdccxxxix, osservata nella specula dell' istituto
  delle scienze di bologna},
\bpublisher{nella stamperia di Lelio dalla Volpe},
\blocation{Bologna}.
\end{bbook}
\endbibitem

\bibitem[\protect\citeauthoryear{{Zanotti}}{1742}]{Zanotti1742}
\begin{bbook}
\bauthor{\bsnm{{Zanotti}}, \binits{E.}}:
\byear{1742},
\bbtitle{Osservazioni sopra la cometa dell' anno mdccxxxxii},
\bpublisher{nella stamperia di Lelio dalla Volpe},
\blocation{Bologna}.
\end{bbook}
\endbibitem

\bibitem[\protect\citeauthoryear{{Zanotti}}{1744}]{Zanotti1744}
\begin{bbook}
\bauthor{\bsnm{{Zanotti}}, \binits{E.}}:
\byear{1744},
\bbtitle{Osservazioni sopra la cometa dell' anno mdccxliv},
\bpublisher{Nella stamperia di Lelio dalla Volpe},
\blocation{Bologna}.
\end{bbook}
\endbibitem

\bibitem[\protect\citeauthoryear{{Zolotova} and
  {Ponyavin}}{2016}]{2016SoPh..291.2869Z}
\begin{barticle}
\bauthor{\bsnm{{Zolotova}}, \binits{N.V.}},
\bauthor{\bsnm{{Ponyavin}}, \binits{D.I.}}:
\byear{2016},
\batitle{{How Deep Was the Maunder Minimum?}}
\bjtitle{\solphys}
\bvolume{291},
\bfpage{2869}.
\doiurl{10.1007/s11207-016-0908-z}.
\adsurl{2016SoPh..291.2869Z}.
\end{barticle}
\endbibitem

\end{thebibliography}
\end{article} 
\end{document}